\documentclass[reprint,prb,amsmath,amssymb,aps]{revtex4-1}

\usepackage{graphicx}
\usepackage{dcolumn}
\usepackage{bm}
\usepackage{longtable}

\begin{document}

\title{A density functional theory study of FeAs comparing LDA+U, GGA+U and hybrid functionals}

\author{Sin\'{e}ad M. Griffin}
\email{sineadmgriffin@umail.ucsb.edu}
\homepage{http://sites.google.com/sineadv0}
\affiliation{Materials Theory, ETH Zurich,
Wolfgang-Pauli-Strasse 27, CH-8093 Zurich, Switzerland} 

\author{Nicola A. Spaldin}
\email{nicola.spaldin@mat.ethz.ch}
\homepage{http://www.theory.mat.ethz.ch}
\affiliation{Materials Theory, ETH Zurich,
Wolfgang-Pauli-Strasse 27, CH-8093 Zurich, Switzerland} 

\date{\today}

\begin{abstract}
We use density functional theory within the local density approximation (LDA), LDA+U, generalised gradient approximation (GGA), GGA+U, and hybrid-functional methods to calculate the properties of iron monoarsenide. FeAs, which forms in the MnP structure, is of current interest for potential spintronic applications as well as being the parent compound for the newly-identified pnictide superconductors.  We compare the calculated structural, magnetic and electronic properties obtained using the different functionals to each other and to experiment, and investigate the origin of a recently-reported magnetic spiral. Our results indicate the appropriateness or otherwise of the various functionals for describing FeAs and the related Fe-pnictide superconductors.
\end{abstract}


\maketitle

\section{Introduction}

The recent discovery of fairly high-temperature and possibly unconventional superconductivity in doped LaOFeAs provoked intensive investigation into Fe-pnictogen containing compounds\cite{Hosono:2008, Norman:2008, Wang:2008, Yin:2008}. Superconductivity has since been found in many related materials, all of which share the common structural motif of a Fe-pnictogen sandwich, with the surrounding layers acting as spacers and
charge donors/acceptors. Experiment and theory both point to the FeAs layers as the locus for understanding superconductivity\cite{Hosonoreview:2009, Eremin:2008, Maier:2008, Yildrim:2008}. Although the pairing 
mechanism is not yet known, many structural and electronic properties of the superconducting family have been characterized. The phase diagram as a function
of doping has been intensely investigated and many competing structural and magnetic orderings identified\cite{Dai:2008}; of particular interest is the antiferromagnetic phase which lies next to the
 superconducting region as a function of both doping and pressure. Several correlations have been suggested between the superconducting transition temperature and structural parameters of the FeAs layer such as the pnictogen height and the tetrahedral bond angle, showing the importance of understanding the structure-property relationships in this class of materials\cite{Arita:2009, Lee:2008}.

FeAs-based compounds had previously been widely studied for their potential ``spintronic'' functionalities\cite{Prinz:1990, Munekata:1989, Ohno:1996, Shirai:2001, Rahman:2006}. These are magnetic semiconducting materials in which the charge degree of freedom is augmented by a spin component, making it desirable for magnetic storage.

Density functional theory (DFT) has been utilised successfully to address many questions regarding the properties of the iron-pnicitide superconductors and the related Fe-As spintronic compounds\cite{Singh:2008, Cao:2008, Dong:2008, Griffin/Spaldin:2012}. The ground-state structural and magnetic properties have been correctly reproduced with DFT, provided due care is taken with the choice of exchange-correlational functional. 
 Band structure calculations have confirmed the semimetallic nature of the compounds,
 finding hole and electron pockets at the Fermi level, suggesting that the pairing mechanism may be 
related to Fermi surface nesting. DFT has also confirmed the ground state as striped antiferromagnetic ordering for all of the parent compounds of the superconducting pnictides\cite{Kroll:2008}.

Despite its many achievements, there have been some problems with using DFT to model the Fe-pnictide materials. The choice of the exchange-correlation functional poses particular difficulties in these compounds since they lie between the weakly- and strongly-correlated limits. Studies comparing the local density approximation (LDA) with the generalized gradient approximation (GGA) have concluded that, while GGA gives better structures, the two give similar results for magnetic and electronic properties\cite{Mazin:2008}. The theoretically-determined magnetic moment on the Fe sites is greatly overestimated within both the LDA and GGA, 
which is an unusual failing of DFT\cite{Sushko:2008}. In addition, the pnictogen height with respect to the Fe planes is consistently underestimated compared to experiment\cite{Singh:2009}. For this reason, many electronic structure calculations are now carried out using the experimentally-determined structure.

While the Fe environment is different in iron monoarsenide from that in the Fe-pnictide superconductors (the former has a bulk octahedral network while the latter consists of tetrahedrally coordinated layers), both compounds lie at the boundary between itinerant and localized magnetism. Since many of the computational issues for the pnictides are related to this dichotomy -- DFT exchange-correlation functionals exist which successfully describe localized-moment insulators and simple metals, but how to best treat the meeting of these extremes is an ongoing question -- a methodology that describes well the structural and magnetic features of bulk FeAs will likely also be appropriate for the more complex Fe-pnictide superconductors.

In this work we perform a systematic investigation of the effects of the choice of DFT exchange-correlation functional on the calculated properties of the parent iron pnictide compound FeAs. We study the bulk ground state, MnP-type FeAs, and calculate the crystal structure, magnetic
properties and electronic structure using both the well-established LDA and GGA functionals and their ``+U'' extensions, as well as the recently introduced hybrid functional\cite{HSE}. Our goal is to identify the most appropriate functional for describing FeAs, and in turn the pnictide superconductors, and to understand the fundamental physics underlying the choice.

\section{Existing Literature}
The ground state of bulk FeAs is the orthorhombic MnP-type structure which it adopts up to its melting temperature of 1300 K (Fig. 1). The primitive unit cell contains eight atoms, with the Fe ions, coordinated by distorted As octahedra, forming zig-zag chains along the $a$-axis. 

The space group of bulk FeAs is still controversial. The first experimental characterization of the structural and magnetic properties was performed in 1969 by Selte et al.\cite{Selte:1969}  using X-ray diffraction. They found the ground state to have $Pnma$ symmetry, adopting the same structure as MnP. Next, Lyman and Prewitt suggested the space group $Pna2_{1}$ based on a comparison of x-ray refinements in both structures\cite{Lyman:1984}. More recently, Rodriguez et al.\cite{Rodriguez:2011} performed both powder and single-crystal neutron diffraction experiments again finding $Pnma$ symmetry. 

Furthermore, the magnetic properites of this metallic antiferromagnetic are not fully understood. The first magnetic study using powder neutron diffraction indicated a simple incommensurate spin spiral of wavevector \textbf{q}=0.375\textbf{c*} with the moments lying in the $ab-$plane\cite{Selte:1972}. This was later disputed by transport experiments which revealed highly anistropic magnetic properties, where the susceptibility along the $a$ and $b$ axis was found to differ greatly\cite{Segawa:2009}. More recently, Rodriguez et al.\cite{Rodriguez:2011} performed single-crystal neutron-diffraction experiments to elucidate the nature of the magnetic ordering. They proposed incommensurate modulated magnetism, however were unable to distinguish between the previously proposed simple spiral structure\cite{Selte:1972}, or a collinear spin-density wave structure. They confirmed the highly anistropic magnetism obtaining ~15\% greater spin polarization in the $b$-plane compared to the $a$-direction.  Further confirmation of the anistropy also came from M\"{o}ssbauer measurements from Blachowski et al.\cite{Blachowski:2014}. However, neither Refs. 28 or 30 were able to give a precise conclusion for the spin structure. 

Very little theoretical work has been carried out on bulk MnP-type FeAs. First principles calculations performed by Parker and Mazin\cite{Parker/Mazin:2011} confirmed the antiferromagnetic ground-state magnetic ordering. However, unlike the Fe-based superconductors, this could not be explained by Fermi-surface nesting. They also found a 3-dimensional Fermi surface, lending support to the anistropic magnetic behavior from experiments.

\section{Calculation Details}

We performed density functional calculations as implemented in the Vienna ab initio Simulation Package (VASP)\cite{VASP1, VASP2} and wavefunctions expanded in plane waves to an energy cutoff of 500 eV. We used the projector augmented wave (PAW) methods for the electron-core interactions with Fe(3d, 4s) and As(4s, 4p) shells treated as valence. Since MnP-type FeAs is close to semimetallic, a dense Brillouin zone
 sampling scheme of 10x10x10 Monkhorst-Pack grid was used\cite{Monkhorst_Pack}. The internal coordinates were relaxed in all cases until the Hellmann-Feynmann
 forces were less than 1 meV/\AA\  on each atom. Equations of state were fitted with the Murnaghan-Birch equation\cite{EOS:1983}.

Comparisons were made between spin-polarised LDA, PBE\cite{PBE1, PBE2}, LDA+U and GGA+U, in addition to hybrid functional calculations.
DFT+U calculations were performed in the Dudarev scheme with an effective $U_{eff}=U-J$
where $U$ represents the electron-electron correlation term and $J$ is the electron exchange energy\cite{Dudarev}. The value of $U_{eff}$ was varied between -2 eV and 4 eV. In the hybrid functional calculations, the standard HSE functional with 75\% PBE and 25\% exact Hartree-Fock exchange was used\cite{HSE}. 

\section{Results}

\subsection{Ground State Structure}
First, we address the question of the ground state structure by comparing the calculated energies of the $Pnma$ and $Pna2_{1}$ structures. In both cases, the initial structure was taken from experiments and then a full relaxation of the lattice parameters and internal coordinates was performed with symmetry constrained. The ground state structure was found to be the $Pnma$ structure, with an energy $4$ meV per formula unit (f.u.) and $17$ meV/f.u. lower than the $Pna2_{1}$ for the LDA and GGA calculations respectively. For the rest of this work we focus therefore on the $Pnma$ structure.

\begin{figure}[h]
 \centering
 \includegraphics[width=8cm, keepaspectratio=true]{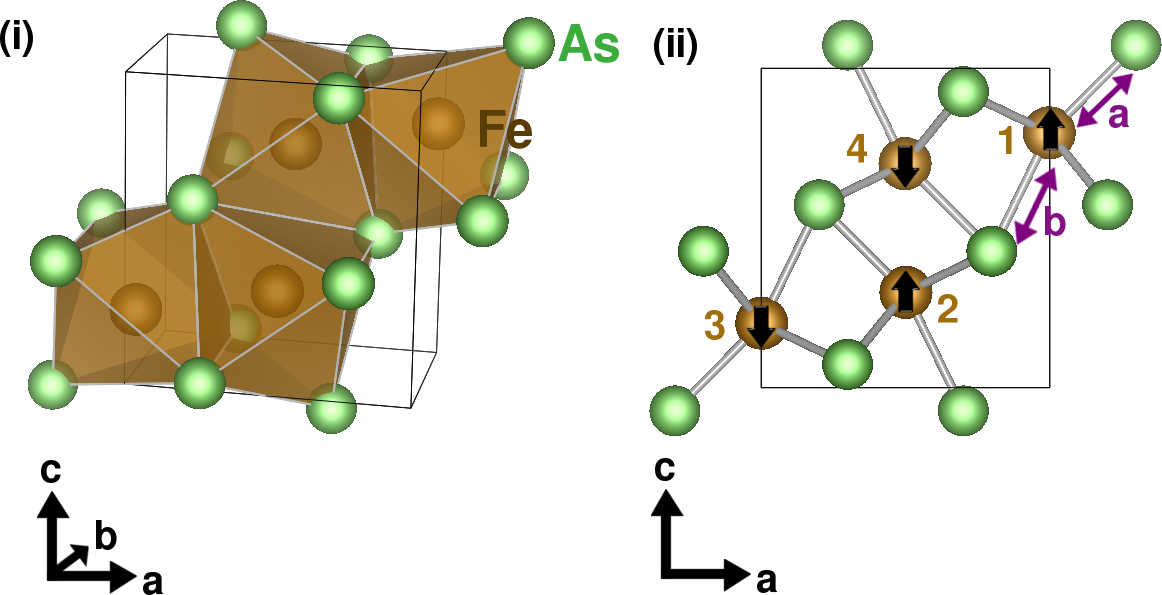}
 \caption{(i) The MnP-type structure of bulk FeAs where each Fe is 6-fold coordinated by As in a network of edge-sharing polyhedra. (ii) View in the [010] direction of the MnP unit cell. The Fe sites are labelled for later discussion of bond lengths, likewise the shortest (a) and longest (b) Fe-As bondlengths are shown. The black arrows indicate the lowest-energy collinear AFM order.}
 \label{Fig. 3.}
\end{figure}

Next we compare our cell parameters and atomic positions calculated with the five considered functionals with experiment (Table 1). All results are for collinear AFM ordering where the nearest neighbours are antiferromagnetically coupled, and the next-nearest are ferromagnetically coupled (Fig. 1 (ii)). We later show this to be the lowest energy commensurate arrangement. We see that LDA greatly 
underestimates the experimental unit cell volume of $110.64$ \AA$^3$ by around 9\%, which is larger than the usual LDA underestimation. However, both GGA and HSE volumes agree well with experiment, with values of $109.3$ \AA$^3$ and $108.4$ \AA$^3$ respectively. The usual behavior of GGA is to overestimate the volume, but here the calculated value is 
~1\% less than experiment. Interestingly the inclusion of the Hartree-Fock exact exchange term does not significantly alter the GGA structure.

\begin{center}
\begin{table*}[ht]
\caption{\label{1} Calculated lattice parameters and atomic fractional coordinates obtained using the LDA, GGA and HSE functionals as well as the experimental values from Ref.[\onlinecite{Selte:1969}]}.
\begin{ruledtabular}
\begin{tabular}{l|c|ccc|cccc|cc|cc}
 & Vol (\AA$^{3}$) & a(\AA) & b(\AA) & c(\AA) & Fe\textit{(x)} & Fe\textit{(z)} & As\textit{(x)} & As\textit{(z)} & Fe-Fe (a) & Fe-Fe (b) & Fe-As (1) & Fe-As (2) \\ \hline
 LDA & 100.33 & 5.313 & 3.194 & 5.912 & 0.0017 & 0.2017 & 0.2000 & 0.5728 & 2.717 & 2.871 & 2.281 & 2.434  \\ 
 GGA & 108.46 & 5.468 & 3.277 & 6.051 & 0.0017 & 0.2016 & 0.2007 & 0.5728 & 2.796 & 2.939 & 2.338 & 2.496\\ 
 HSE+PBE & 108.40 & 5.470 & 3.276 & 6.050 & 0.0019 & 0.2016 & 0.2007 & 0.5727 & 2.797 & 2.938 & 2.338 & 2.495 \\ 
 Exp\cite{Selte:1969} & 110.64 & 5.442 & 3.373 & 6.028 & 0.0027 & 0.1994 & 0.1992 & 0.5773 & 2.788 & 2.937 & 2.347 & 2.516 \\ 
\end{tabular}
\end{ruledtabular}
\end{table*}
\end{center}

In Fig. 2 (upper panel) we show the effect on the calculated volume of adding a U to the GGA functional. Interestingly, the volume increases by over 20\% as U is increased from 0 eV to 4 eV  leading to unphysically large volumes for the higher U values. The volume does not change significantly for a negative U, remaining close to the U=0 eV and experimental value.

The calculated shortest Fe-As bond distance was found to be $2.28$ \AA\ for LDA, $2.34$\AA\  for GGA, and $2.34$ \AA\  for the hybrid functional, compared to the experimental value of $2.35$ \AA. Here again, the LDA performs poorly with a substantional underestimation of the bond distance. GGA, GGA+U (for U$<$1 eV) and the hybrid functional values are very close to the experimental value.  

In conclusion, the hybrid functionals along with GGA and GGA+U for small values of U give the best match to the experimental lattice parameters and internal coordinates.

\subsection{Magnetic Properties}

\begin{figure}
 \includegraphics[width=10cm, keepaspectratio=true]{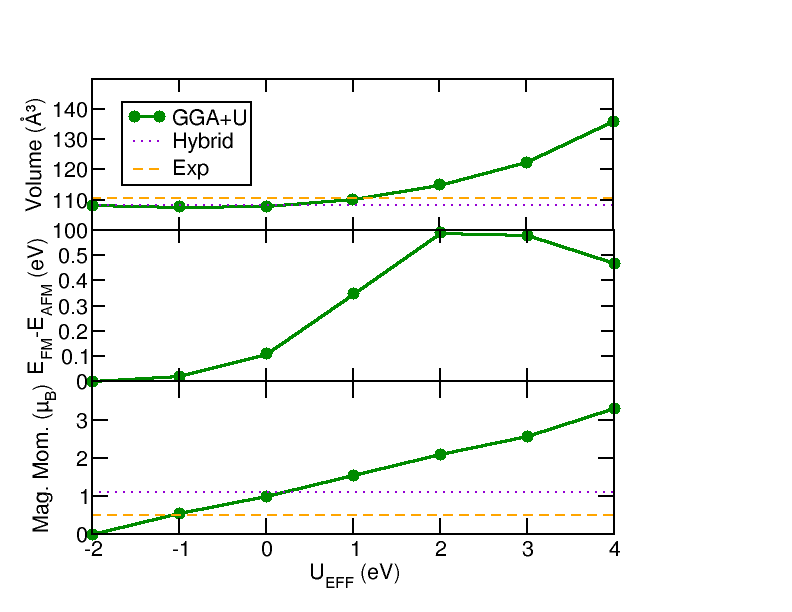}
 \caption{\label{2} Calculated unit cell volume, magnetic energy, and magnetic moment within the GGA+U method as a function of U. The top panel shows the volume of the unit cell containing four formula units; the middle panel shows the energy difference between the ferromagnetic and antiferromagnetic orderings per eight atoms, and the bottom shows the magnetic moment of one Fe site. The experimental values of the Fe magnetic moment and cell volume are indicated by the dashed orange lines. The results using the hybrid functional are shown with the dotted purple lines.}
 \label{Fig. 2}
\end{figure}

Next we compare the calculated magnetic ordering and magnetic moment size obtained with different functionals. To isolate the effects of magnetic ordering, we use the same structure -- that obtained from optimized hybrid functional calculations -- for all calculations. We obtain an antiferromagnetic ground state with nearest-neighbour Fe ions coupled antiferromagnetically, and second-nearest neighbours ferromagnetically coupled, for all functionals and all values of U.

\begin{figure}
 \includegraphics[width=7cm, keepaspectratio=true]{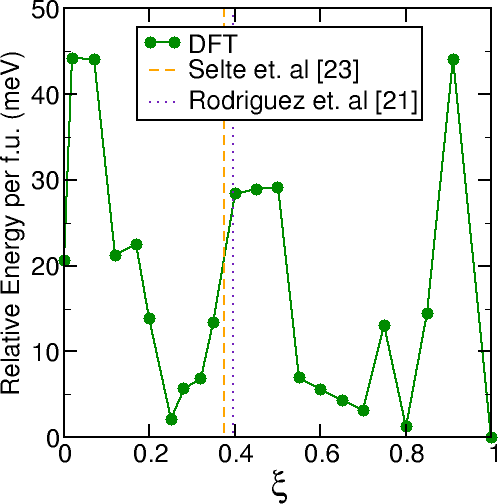}
 \caption{\label{2} Calculated energies for a spiral wavevector propagating along the \textbf{c}-axis. In our units, $\xi=0$ corresponds to FM ordering, and $\xi=1$ to nearest-neighbour AFM. The experimental result of Selte et al. [23] of $\textbf{q}=0.375\times \mathbf{c*}$ is shown by the dashed orange line, and of Rodriquez et al. [21] of $\textbf{q}=0.395\times \mathbf{c*}$ by the dotted purple line.}
 \label{Fig. 6}
\end{figure}

Fig. 2 (middle panel) shows our calculated total energy difference between the FM ordering and the ground-state AFM ordering as a function of U.
Note that increasing the value of U in the GGA+U method increases the relative stabilization energy for AFM. This shows that increasing U moves the system away from any frustration caused by competition between AFM and FM ordering. The bottom panels shows the calculated magnetic moment as a function of U. As expected, the moment steadily increases with higher values of U as a result of the increased localization of the bands.
As in the case of the Fe-pnictide superconductors, all three functionals overestimate the value of the magnetic moment. The deviation scales with U in the GGA+U method, and the experimental value of magnetic moment can be attained for a negative $U_{eff}=-1$eV.

As discussed above, the experimental bulk structure has an incommensurate antiferromagnetic ordering. Such incommensurate magnetic ground states often result either from electronic instabilities such as nesting, or from competition between ferromagnetic and antiferromagnetic exchange interactions. To investigate a possible electronic origin, we perform spin-spiral calculations as implemented in the VASP code. This allows us to impose helimagnetism (a spin spiral in which neighboring spins tilt by a fixed angle with constant amplitude) by modifying the periodic boundary conditions in the system. Spin-orbit coupling is not considered and so the role of the electronic structure is isolated from that of the lattice.

In Fig. 3 we plot our calculated total energy as a function of the propagation wavevector $\xi$. The overall energy minimum is for $\xi=1$, which corresponds to a commensurate antiferromagnetic coupling of the nearest-neighbour Fe ions. The spiral wavevectors reported by Selte et al. and Rodriguez et al. are shown by the vertical orange and purple lines; both occur at higher energies.

We next calculated the magnetocrystalline anistropy energy (MAE) by including spin-orbit coupling and calculating the energy difference for spins aligned along the x, y and z axes. Our calculations show a preference for spins to lie along the x-axis with energy differences of 0.07 meV/f.u. and 0.08 meV/f.u. compared to the y-axis and z-axis respectively.

We conclude, therefore, that the reported incommensurate magnetism is unlikely drive by an electronic instability causing a helimagnetic spiral, and that coupling to the lattice is important when considering the origin of magnetism in this compound.

\subsection{Electronic Properties}

\begin{figure*}
 \centering
 \includegraphics[width=18cm, keepaspectratio=true]{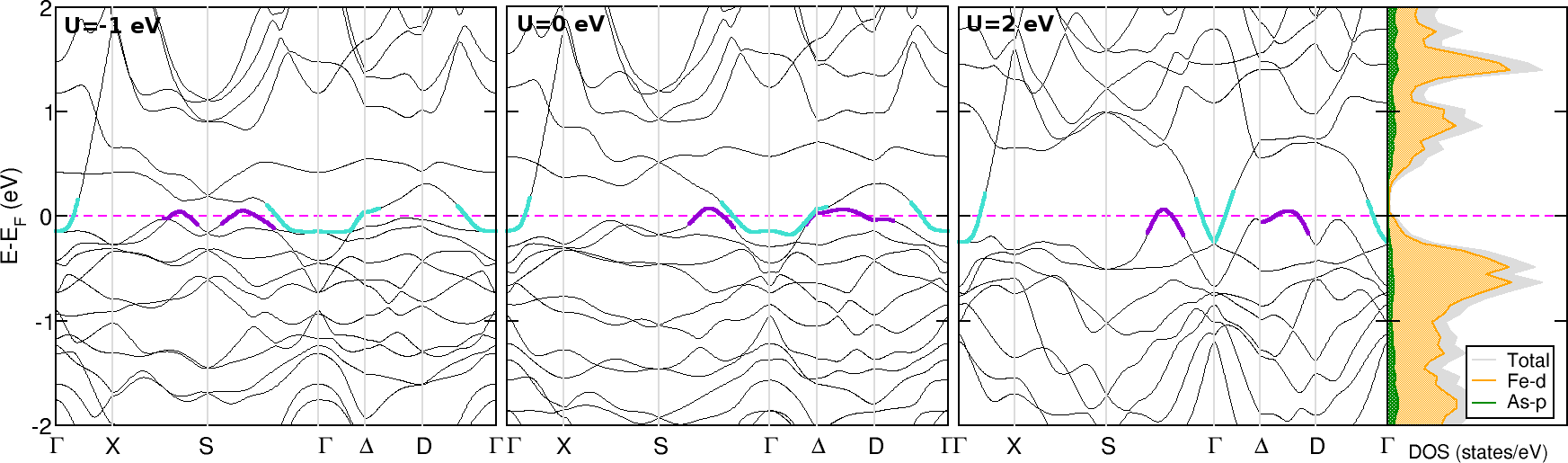}
 \caption{\label{2} Calculated band structures for MnP-type FeAs using the GGA+U method with U=-1,0,2 eV. The density of states for U=2 eV is also shown next to the U=2 eV band structure. The orbitally-projected Fe-\textit{d} states are indicated by the dashed orange line and region, the As-\textit{p} by the shaded green region. In all plots the Fermi level has been set to 0 eV and is marked by the dashed line.}
 \label{Fig. 8}
\end{figure*}

Fig. 4 shows the calculated band structures of FeAs for values of U ranging from -1 eV to 2 eV, again for the optimized structure calculated with the hybrid functional. The density of states for U=0 eV was also calculated and is shown in the figure. In each case the region around the Fermi level is composed of a broad band of Fe-\textit{d} states with a small contribution from As-{p} states. Compared with the U=0 eV case, the negative-U band structure has more delocalized bands, consistent with the lower value of magnetic moment on the Fe-{d} states. Increasing the U value has the expected result of pushing bands away from the Fermi level, increasing any gaps between bands at E$_{F}$. 

However, depsite this general trend of band localization, the main features of the electronic structure do not change significantly on varying the U parameter. In each case we find an electron pocket at the $\Gamma$ point (shown in turquoise in Fig. 4), whose size and dispersion are slightly U-dependent. The curvature of the pocket increases with U giving higher electron effective masses. We also find two hole pockets for each case (marked in purple), however their location in the Brillouin Zone changes as a function of U. Along the S to $\Gamma$ direction, a Fermi surface pocket is present for all U values. However, for U=-1 eV, a second pocket appears in the X to S direction which is not present in the U=0 eV or U=2 eV cases. In the latter two calculations, we instead find the second pocket in the $\Delta$ to D direction. As U is increased, this first X to S pocket as seen in the U=-1 eV case is pushed further away from the Fermi level.
In summary, the character of the bands at E$_{F}$ changes modestly with increasing U, however these subtle changes could have a big effect on Fermi surface nesting since the location of the pockets changes as a function of U.

\section{Discussion}

Perhaps surprisingly for this ostensibly simple ferropnictide compound, we encountered similar problems with using density functional theory to calculate its structural and magnetic properties to those reported for the pnictide superconductors. As in the case of the Fe-pnictide parent compounds, simple LDA calculations do not reproduce the  measured structural parameters, with a 10\% LDA underestimation of the cell volume in this case. We also find that the Fe-As distance is underestimated in the LDA, which is a common failing in the Fe-pnictide literature. However, both the GGA and hybrid functionals give structures that are much closer to experiment. Subsequent addition of a Hubbard-U to the GGA calculations has little effect on the volume for U$<2$eV (including for negative U), but larger U values cause a strong divergence from the experimental volume. The conclusion for structural calculations echoes that previously made for the ferropnictide superconductors; that GGA best reproduces the structural parameters while LDA does extremely badly.

Next we examined the magnetic properties using LDA, LDA+U, GGA, GGA+U and the hybrid functional. The lowest energy magnetic ordering was found to be antiferromagnetic, consistent with the spin spiral structure found in experiments where the nearest neighbour spins are antiferromagnetically coupled. The simple spin spiral as proposed by Selte and later Rodriguez  was not found to be the ground state when helimagnetic calculations were performed. However, magnetocrystalline anistropy calculations compare well with experiment -- there is a preference for the spins to lie along the x-axis. The most likely magnetic ordering from our calculations is some modulated antiferromagnetism as proposed by Rodriguez in which a noncollinear spin-density wave traces out an ellipse (rather than a circle in the case of a simple spiral).

As in the case of the pnictide superconductors, the magnetic moment on the Fe ions is strongly overestimated compared with experiment; for example the calculated moment using hybrid functionals and GGA+U is 2.2 times the experimental value. Spin fluctuations have been proposed to account for this discrepancy between theory and experiment\cite{Johannes:2009}; since DFT is a mean field theory it does not include the results of temporal fluctuations. One possible solution we considered was to venture into a \textit{negative}-Hubbard-U regime. However a physical rationale for doing so is not so obvious, despite its match with experiment in both the magnetic moment and structural properties. 

While spin fluctuations are a plausible cause to the overestimation of the moment in density functional theory, the impact of the covalent bonding between Fe and As in these materials should not be overlooked. This was considered recently for the cuprate class of superconductors, where `missing' neutron intensities were found by reinterpreting the neutron data with the strong covalency of the materials\cite{Walters:2009}. In order to fully understand the magnetism of these ferropnictide materials, the extent to which spin fluctuations and/or covalency influence the magnetic moments must be further examined.

\section{Summary}
We performed a thorough investigation of the effect of exchange-correlation functional (LDA, LDA+U, GGA, GGA+U, HSE) on the calculated structural, magnetic and electronic properties of MnP-type FeAs. The hybrid functional and GGA best reproduced the experimental structures and ground-state magnetic ordering in the collinear limit. As is also found in the Fe-pnictide superconductors, LDA performs poorly for the structural calculations, and all functionals overestimate the Fe magnetic moment by at least a factor of 2. Only a \textit{negative}-U regime correctly reproduces the experimental value of the magnetic moment.

To investigate the observed modulated noncollinear magnetic structure, we performed helimagnetic calculations, and found a simple spin-spiral is not the calculated ground state. This suggests a different origin for the incommensurate magnetism such as the elliptical spin-density wave as proposed by Rodriguez et al.\cite{Rodriguez:2011}.

\bibliography{mnp}

\end{document}